\documentclass[aip,reprint,nofootinbib]{revtex4-1}
\usepackage{amsmath,amssymb}
\usepackage{graphicx}
\usepackage{dcolumn}
\usepackage{bm}
\usepackage[utf8]{inputenc}
\usepackage[T1]{fontenc}
\usepackage{times}
\usepackage{mathptmx}
\usepackage{xcolor}
\usepackage{soul}
\usepackage[colorlinks=true,linkcolor=black,citecolor=black,urlcolor=blue]{hyperref}
\usepackage[normalem]{ulem}
\usepackage{xspace}
\usepackage{array}

\newcolumntype{P}[1]{>{\centering\arraybackslash}p{#1}}
\begin{document}

\title[]{Discrete-time Kuramoto model with phase lag: Linear stability analysis and onset of synchronization}
\author{Prashant M. Gade}
\affiliation{Ramniranjan Jhunjhunwala College of Arts, Science and Commerce, Ghatkopar West, Mumbai, 400086, India}
\author{Shamik Gupta}
\affiliation{Department of Theoretical Physics, Tata Institute of Fundamental Research, Homi Bhabha Road, Mumbai 400005, India}
\email{prashant.m.gade@gmail.com}
\email{shamik.gupta@theory.tifr.res.in}
\date{\today}

\begin{abstract}
We investigate the discrete-time version of the Kuramoto model with phase lag, which comprises globally-coupled phase oscillators of distributed frequencies that are evolving under a nonlinear map. In 
the continuum limit of an infinite number of oscillators ($N\to \infty$), we derive the exact Frobenius-Perron equation for the time evolution of the single-oscillator probability density, and 
study linear stability of the incoherent state. Instability signals onset of synchronization. The corresponding 
synchronization threshold is obtained analytically for the case of a Lorentzian  distribution of the oscillator frequencies. The threshold differs from that of the continuous-time Kuramoto model, reflecting the fundamentally different stability conditions for discrete-time maps and continuous-time flows. Beyond synchronization threshold, we observe several interesting nonlinear phenomena: Unlike the classical Kuramoto model, the discrete-time version exhibits periodic and chaotic states.  Numerical simulations of the finite-$N$ system confirm the analytical prediction for the synchronization threshold, while highlighting breakdown of the celebrated Ott-Antonsen ansatz invoked to conveniently study the continuous-time Kuramoto model in terms of a low-dimensional description. 
\end{abstract}
\maketitle

\begin{quotation}
The Kuramoto model has provided a cornerstone for understanding synchronization in large populations of coupled oscillators, yet its discrete-time counterpart remains less explored. We show that replacing continuous-time evolution by a nonlinear map representing the discrete-time dynamics fundamentally changes the synchronization physics. For an oscillator frequency distribution that is Lorentzian, we derive an exact Frobenius--Perron description and obtain an analytical synchronization threshold that differs from the continuous-time result. We also demonstrate synchronization under repulsive coupling between the oscillators, even when the phase lag in the dynamics equals $\pi$ and synchronization is not possible in the continuous-time version. Beyond the synchronization transition, the discrete dynamics supports collective period-two and period-three states and signatures of chaos, while the Ott--Antonsen manifold is no longer preserved. Our results demonstrate that discrete-time synchronization constitutes a distinct dynamical regime and provide an analytical foundation for its systematic study.
\end{quotation}

Keywords: Spontaneous synchronization, Kuramoto model, Discrete-time evolution, Linear stability analysis

\section{Introduction}

The phenomenon of spontaneous synchronization in large ensembles of interacting oscillators is ubiquitous across physical, biological, and technological systems~\cite{Pikovskybook}. A paradigmatic framework for understanding such collective behavior is the Kuramoto model, which describes a population of phase oscillators coupled through a mean-field interaction. Since its inception, the model has been extensively studied and has found applications ranging from neural dynamics and power grids to chemical and biological oscillations~\cite{kuramoto1984chemical,Strogatz2000,kuramotormp,gupta2014kuramoto,Rodrigues2016,gupta2018statistical}.

A major breakthrough in the analytical treatment of the Kuramoto model was achieved through the introduction of the Ott–Antonsen (OA) ansatz, which enables an exact reduction of the infinite-dimensional dynamics in the continuum limit to a low-dimensional manifold governing the evolution of the macroscopic order parameter~\cite{ott2008low, ott2009long}. This reduction has led to significant advances in understanding synchronization transitions, bifurcations, and collective dynamics in a wide class of coupled oscillator systems.

Most studies of the Kuramoto model and its variants have focused on continuous-time dynamics. However, discrete-time formulations arise naturally in a variety of contexts, including periodically-driven systems, stroboscopic observations, and iterative maps of phase oscillators, as also in digital control and sampled-data networks. Discrete-time oscillator models also arise naturally as numerical integrators of the corresponding continuous-time dynamics, raising the question of whether synchronization thresholds are modified by time discretization itself.  Despite their practical relevance, discrete-time versions of the Kuramoto model remain comparatively less explored, particularly from an analytical standpoint.
In discrete time, the dynamics is governed by nonlinear maps rather than differential equations, and the associated stability criteria and bifurcation structures can differ fundamentally from their continuous-time counterparts. In the context of synchronization of phase oscillators, M\"{o}bius maps have previously been introduced as effective discrete-time models that reproduce several collective phenomena observed in continuous-time Kuramoto-type systems~\cite{pikovsky}, and have been presented as computationally-efficient model of the synchronization transition. In contrast, the present work derives an analytical stability threshold for the incoherent state of a discrete-time Kuramoto model with phase lag and investigates the collective states that emerge beyond this threshold. Our work starts from a simpler, more natural discrete-time Kuramoto model defined directly at the oscillator level and analytically derives the synchronisation threshold, while revealing nontrivial dynamical features beyond the onset of synchronisation.  

Collective dynamics in globally coupled maps and oscillator populations has been studied extensively. Globally coupled circle maps exhibit transitions between coherent, ordered, partially ordered, and turbulent regimes, including chaotic itinerancy \cite{Kaneko1991}. Clustering bifurcations have also been investigated in globally coupled map lattices, where clusters correspond to groups of elements sharing identical or related dynamical states \cite{XieHu1997}. Pikovsky and Kurths \cite{PikovskyKurths1994} examined whether globally coupled maps violate the law of large numbers, highlighting the issue of collective behavior in the large-system limit. In globally coupled oscillator populations, Nakagawa and Kuramoto \cite{NakagawaKuramoto1993} demonstrated collective chaos in identical complex Ginzburg–Landau type oscillators, while Clusella and Politi \cite{ClusellaPoliti2021} investigated irregular collective dynamics in identical Kuramoto–Daido phase oscillators. These studies illustrate different aspects of collective behavior in systems of globally coupled elements.

The present work focuses on the macroscopic dynamics of a heterogeneous population of oscillators evolving in discrete time. Specifically, in this work, we investigate a discrete-time Kuramoto model with a phase lag. In the continuum limit of an infinite number of oscillators and with a Lorentzian distribution of the oscillator frequencies, we derive the exact 
Frobenius-Perron equation for the single-oscillator density and perform a linear stability 
analysis of the incoherent state. The synchronization threshold obtained 
analytically is verified through direct numerical simulations. This threshold differs from that of the continuous-time Kuramoto model, reflecting the fundamentally-different stability conditions for discrete-time maps and continuous-time flows. A natural question is whether the OA ansatz can be extended to the discrete-time case. The OA ansatz requires that the Fourier coefficients of the single-oscillator density 
satisfy $\hat f_m(\omega,n) = [\alpha(\omega,n)]^m$ for all $m$, with $\alpha(\omega,n)$ analytic in the lower-half complex-$\omega$ plane. For discrete-time dynamics, however, we show that the dynamics does not 
generally preserve these conditions.  

\section{The model}
The discrete-time Kuramoto model comprises $N$ globally-coupled phase-only oscillators. Denoting by $\theta_i^{(n)} \in [0,2\pi)$ the phase of the $i$th oscillator, $i=1,2,\ldots,N$, at discrete time $n=0,1,2,\ldots$, the phase update takes place according to the dynamics  
\begin{align}
\theta_{i}^{(n+1)}=\theta_{i}^{(n)}+\omega_i+K r_n \sin(\psi_n - \theta_i^{(n)} - \beta)\quad (\mathrm{mod}\ 2\pi),
\label{eq:EOM}
\end{align}
where $K>0$ is the coupling constant, and $\beta$ is the phase-lag parameter. For $0<\beta <\pi/2$, the coupling between the oscillators is attractive, while for $\pi/2<\beta < \pi$, it is repulsive. The complex order parameter, characterizing macroscopic order in the system in terms of phase synchrony, is defined as
\begin{align}
Z_n \equiv r_n e^{i\psi_n}=\frac{1}{N}\sum_{j=1}^N e^{i\theta_j^{(n)}}.
\label{eq:Z-defn}
\end{align}
In Eq.~\eqref{eq:Z-defn}, the quantity $0 \le r_n \le 1$ is a measure of the amount of phase synchrony present in the system at time $n$, while $\psi_n \in [0,2\pi)$ denotes the corresponding average phase. Full synchrony (respectively, incoherence) is characterized by the value $r_n=1$ (respectively, the value $r_n=0$). The natural frequencies $\omega_i \in (-\infty,\infty)$ in Eq.~\eqref{eq:EOM} are quenched-disordered random variables that we choose to be sampled independently from a Lorentzian distribution 
\begin{align}
g(\omega)=\frac{\gamma}{\pi[(\omega-\omega_0)^2+\gamma^2]}.
\label{eq:lorentzian}
\end{align}
Here, the parameter $\gamma>0$ denotes the half width at half maximum (HWHM) of the distribution, while $\omega_0$ denotes the location of the peak of the distribution. The parameters $\omega_i$, $K$, $\gamma$, $\omega_0$ are all dimensionless.

The dynamics~\eqref{eq:EOM} has $O(2)$ symmetry, whereby the dynamics remains invariant when all the phases are rotated by the same angle. Then, effecting the transformation $\omega_i \to \omega_i-\omega_0$ and $\theta_i^{(n)} \to \theta_i^{(n)}-\omega_0 n ~\forall~i$ (tantamount to viewing the dynamics in a rotating frame), one may get rid of the effect of the quantity $\omega_0$ from the dynamics. Consequently, from now on, we will consider Eq.~\eqref{eq:EOM} in which $\omega_0$ has been set to zero. The discrete dynamics defines the nonlinear map
\begin{align}
\theta_i^{(n+1)} = T_n(\theta_i^{(n)}),
\label{eq:Tmap}
\end{align}
with
\begin{align}
T_n(\theta) \equiv \theta + \omega + K r_n \sin(\psi_n -\theta - \beta).
\label{eq:Ttheta-defn}
\end{align}

In the limit $N\to\infty$, it is convenient to describe the system in terms of a single-oscillator probability density $f(\theta,\omega,n)$, which denotes the fraction of oscillators with natural frequency $\omega$ that have phase 
$\theta$ at discrete time $n$. The density satisfies the normalization
\begin{align}
\int_0^{2\pi}d\theta~f(\theta,\omega,n)= g(\omega)~\forall~n,
\label{eq:normalization}
\end{align}
while the order parameter \eqref{eq:Z-defn} may be written as
\begin{align}
Z_n=\int_{-\infty}^\infty d\omega \int_0^{2\pi} d\theta~e^{i\theta}f(\theta,\omega,n).
\label{eq:Z-defn-1}
\end{align}

Because the dynamics \eqref{eq:Tmap} is deterministic, the evolution of $f$ is governed by 
the Frobenius-Perron (FP) operator associated with the map $T_n$. Consider the set of oscillators 
with frequency $\omega$ that at time $n+1$ have phases in $[\theta, \theta + d\theta]$. 
These must have come from phases $[\theta', \theta' + d\theta']$ that these oscillators have had at time $n$,  such that 
$\theta = T_n(\theta')$. Conservation of probability for each $\omega$ gives $f(\theta,\omega,n+1)d\theta = f(\theta',\omega,n)d\theta'$. Using $\theta = T_n(\theta')$ and $d\theta = |T_n'(\theta')| d\theta'$, we obtain the FP equation
\begin{align}
f(\theta,\omega,n+1)=\frac{f(T_n^{-1}(\theta),\omega,n)}{\left|T_n'(T_n^{-1}(\theta))\right|},
\label{eq:Frobenius}
\end{align}
with $T_n'(\theta)= 1 - K r_n \cos(\psi_n - \theta - \beta)$. For the map to be locally invertible, we require $K r_n < 1$. 

The density $f$ being $2\pi$-periodic in $\theta$ admits the Fourier expansion
\begin{align}
f(\theta,\omega,n)=\frac{g(\omega)}{2\pi}\left[1+\sum_{m=1}^{\infty} \hat f_m(\omega,n) e^{im\theta}+ \text{c.c.}\right],
\label{eq:fourier-expansion}
\end{align}
where c.c. denotes complex conjugation. Moreover, $f$ being real, we have $\hat f_{-m}=\hat f_m^*$, where star denotes complex conjugation. The prefactor $g(\omega)/(2\pi)$ ensures that Eq.~\eqref{eq:normalization} is satisfied. From Eq.~\eqref{eq:Z-defn-1}, we have
\begin{align}
Z_n = \int_{-\infty}^\infty d\omega~g(\omega) \hat f_1^*(\omega,n),
\label{eq:Z-an}
\end{align}
and so the order parameter $Z_n$ is completely determined by the first Fourier coefficient.

\section{Linearized dynamics and synchronization threshold}
Now, by definition,
\begin{align}
\hat f_1(\omega,n)= \frac{1}{g(\omega)}\int_0^{2\pi}d\theta~ f(\theta,\omega,n) e^{-i\theta},
\label{eq:fourier-coeff-def}
\end{align}
and so, using Eq.~\eqref{eq:Frobenius}, with the change of variables $\theta=T_n(\theta')$, we get straightforwardly that 
\begin{align}
\hat f_1(\omega,n+1)=\frac{1}{g(\omega)}\int_0^{2\pi}   d\theta'~f(\theta',\omega,n) e^{-i T_n(\theta')}.
\label{eq:hierarchy}
\end{align}
Using the expression for $T_n(\theta)$ in Eq.~\eqref{eq:Ttheta-defn}, the exponential in the integrand becomes
\begin{equation}
e^{-i T_n(\theta')}=e^{-i(\theta'+\omega)}\left[1 + i K r_n \sin(\theta' - \psi_n + \beta) + \mathcal{O}(r_n^2)\right].
\end{equation}
Equation~\eqref{eq:hierarchy} then yields to linear order in $r_n$ that
\begin{align}
\hat f_1(\omega,n+1)&=e^{-i\omega}\hat f_1(\omega,n)\nonumber\\
&+\frac{i K r_n}{2\pi} e^{-i\omega}
\int_0^{2\pi} d\theta' \, e^{-i\theta'} \sin(\theta' - \psi_n+ \beta).
\label{eq:hierarchy1}
\end{align}
The  trigonometric integral is elementary:
\begin{equation}
\frac{1}{2\pi}\int_0^{2\pi} d\theta' \, e^{-i\theta'} \sin(\theta' - \psi_n + \beta)=-\frac{i}{2} e^{-i\psi_n} e^{i\beta},
\end{equation}
so that the second term on the right hand side of Eq.~\eqref{eq:hierarchy1} becomes 
\begin{align}
i K r_n e^{-i\omega} \left( -\frac{i}{2} e^{-i\psi_n} e^{i\beta} \right)=\frac{K}{2} e^{i\beta} e^{-i\omega} Z_n^*,
\end{align}
yielding the linearized evolution of the first Fourier mode in compact form as
\begin{equation}
\hat f_1(\omega,n+1)=e^{-i\omega}\hat f_1(\omega,n)+
\frac{K}{2} e^{i\beta} e^{-i\omega} Z_n^*.
\label{eq:linearized_f1_simple}
\end{equation}

To proceed, we assume that $\hat f_1(\omega,n)$ is analytic
in the lower-half complex-$\omega$ plane.
Equation~\eqref{eq:linearized_f1_simple} preserves this property,
since the evolution involves only multiplication by the entire
function $e^{-i\omega}$ and an analytic inhomogeneous term.
Furthermore, $e^{-i\omega}$ decays exponentially on large
semicircles in the lower-half complex-$\omega$ plane, and
$g(\omega)=O(|\omega|^{-2})$ as $|\omega|\to\infty$. Hence, 
Eq.~\eqref{eq:Z-an} may be converted into a contour integral, with the contour running along the real-$\omega$ axis and is closed in the lower-half plane,
so that only the pole of the integrand at $\omega=-i\gamma$ contributes.
Unlike the OA ansatz that requires additionally that $\hat f_m(\omega,n)=[\hat f_1(\omega,n)]^m$, our assumption concerns only the first Fourier mode and imposes no algebraic relation among higher Fourier modes.

\subsection{Synchronization threshold}
On the basis of the foregoing, application of the Cauchy's residue theorem to Eq.~\eqref{eq:Z-an} gives
\begin{align}
Z_n=\hat f_1^{*}(-i\gamma,n).
\label{eq:Zn-pole}
\end{align}
Evaluating Eq.~\eqref{eq:linearized_f1_simple} at $\omega=-i\gamma$ gives
\begin{align}
\hat f_1(-i\gamma,n+1)=e^{-\gamma}\hat f_1(-i\gamma,n)+
\frac{K}{2}e^{i\beta}e^{-\gamma}Z_n^{*}.
\end{align}
Taking the complex conjugate yields
\begin{align}
\hat f_1^{*}(-i\gamma,n+1)=e^{-\gamma}\hat f_1^{*}(-i\gamma,n)+\frac{K}{2}e^{-i\beta}e^{-\gamma}Z_n.
\end{align}
Using Eq.~\eqref{eq:Zn-pole}, we finally obtain
\begin{align}
Z_{n+1}=e^{-\gamma}\left(1+\frac{K}{2}e^{-i\beta}\right)Z_n,
\label{eq:linear-map-final}
\end{align}
which defines a linear map:
\begin{align}
Z_{n+1}=\lambda Z_n;~\lambda=e^{-\gamma}\left(1+\frac{K}{2}e^{-i\beta}\right).
\end{align}
The incoherent state $Z_n=0$ is a fixed point of the map, which loses stability when $|\lambda| = 1$; this gives the  threshold or the critical coupling for the onset of synchronization as
\begin{align}
K_c = 2\left(\sqrt{e^{2\gamma} - \sin^2\beta} -\cos\beta\right).
\label{eq:kc-final}
\end{align}
For special values of the phase lag, Eq.~(\ref{eq:kc-final})
reduces to simple expressions: For $\beta=0$, one obtains
$K_c = 2(e^\gamma-1)$, while for $\beta=\pi/2$, we have $
K_c = 2\sqrt{e^{2\gamma}-1}$. Finally, for $\beta=\pi$, the threshold becomes $K_c = 2(e^\gamma+1)$. The smallest threshold occurs at $\beta=0$, corresponding to purely attractive coupling, while $\beta=\pi$ requires the largest coupling to destabilize the incoherent state. 

The continuous-time limit of Eq.~\eqref{eq:EOM} corresponds to considering $t=n\Delta t$, where $t$ is the continuous time, and $0<\Delta t \ll 1$ is the time step. Accordingly, we need to scale parameters such that $\gamma \to \Gamma=\gamma /\Delta t$, and $K \to \mathcal{K} \equiv K/\Delta t$, and consider the limit $\gamma \to 0$ and $K \to 0$ as $\Delta t \to 0$, so that the quantities $\Gamma$ and $\mathcal{K}$ remain finite. The continuous-time synchronization threshold is obtained from Eq.~\eqref{eq:kc-final} for $\cos \beta>0$, i.e., for $0 \le \beta < \pi/2$, as 
\begin{align}
\mathcal{K}_c=\frac{2\Gamma}{\cos \beta}.
\label{eq:Kc-continuous}
\end{align}
In particular, for $\beta=0$, one obtains $\mathcal{K}_c= 2\Gamma$, which recovers the well-known continuous-time synchronization threshold for a Lorentzian distribution with HWHM equal to $\Gamma$, see Ref.~\cite{Strogatz2000}. For $\beta \ne 0$, the result above matches with the one obtained in Ref.~\cite{Sakaguchi1986}. Note that for the continuous-time version, no synchronization is possible for $\beta \ge \pi/2$.  

To verify the analytical prediction in Eq.~\eqref{eq:kc-final} for the synchronization threshold, we perform direct numerical simulation of the discrete-time model, Eq.~\eqref{eq:EOM}. We consider the initial phases to be drawn uniformly and independently in $[0,2\pi)$. The system is iterated for typically $1.5\times 10^4$ time steps to discard transients, and the order parameter magnitude $r_n$ is then averaged over the next $\sim 10^4$ time steps.

\begin{figure}
\includegraphics[width=\columnwidth]{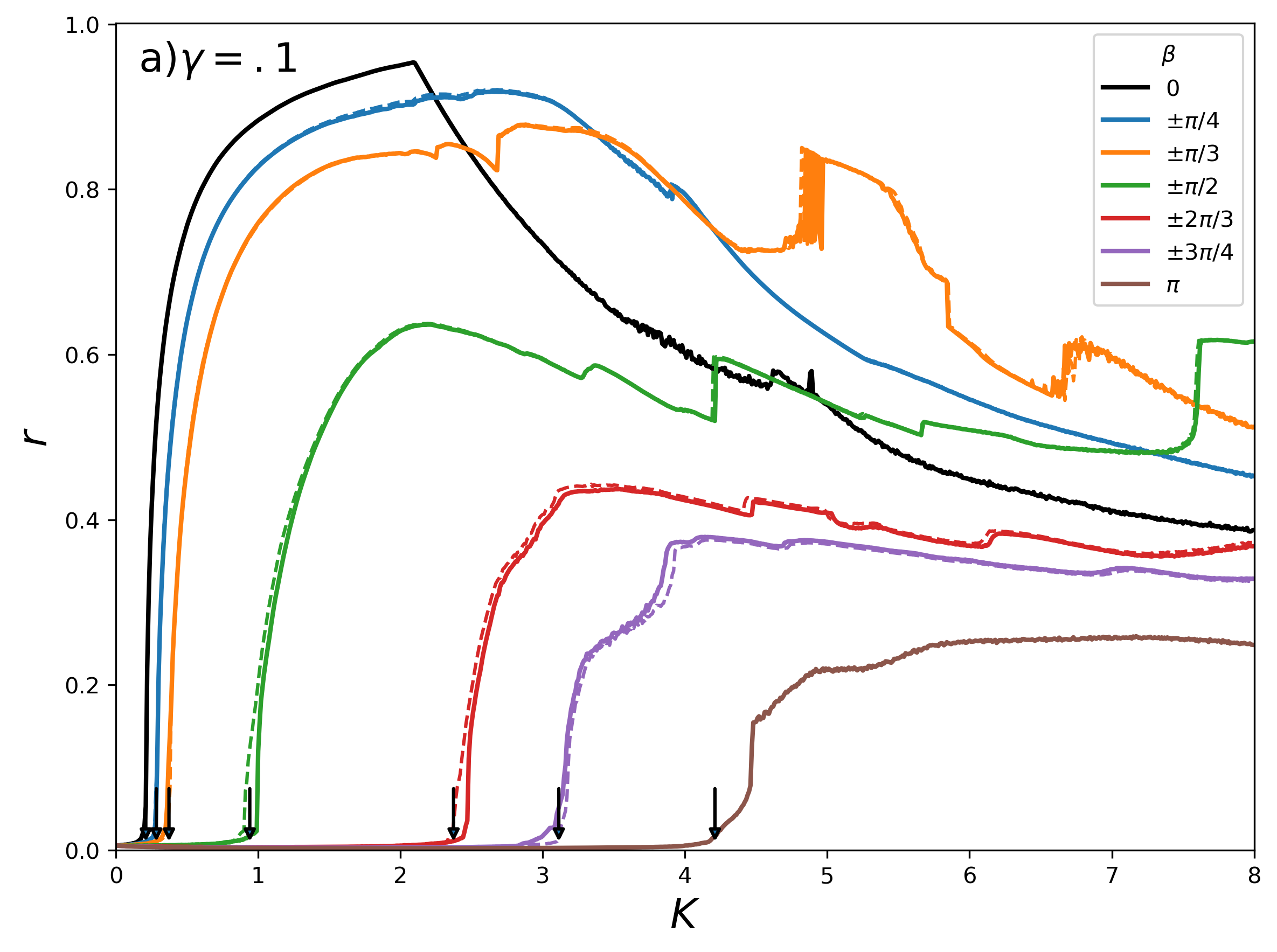}
\includegraphics[width=\columnwidth]{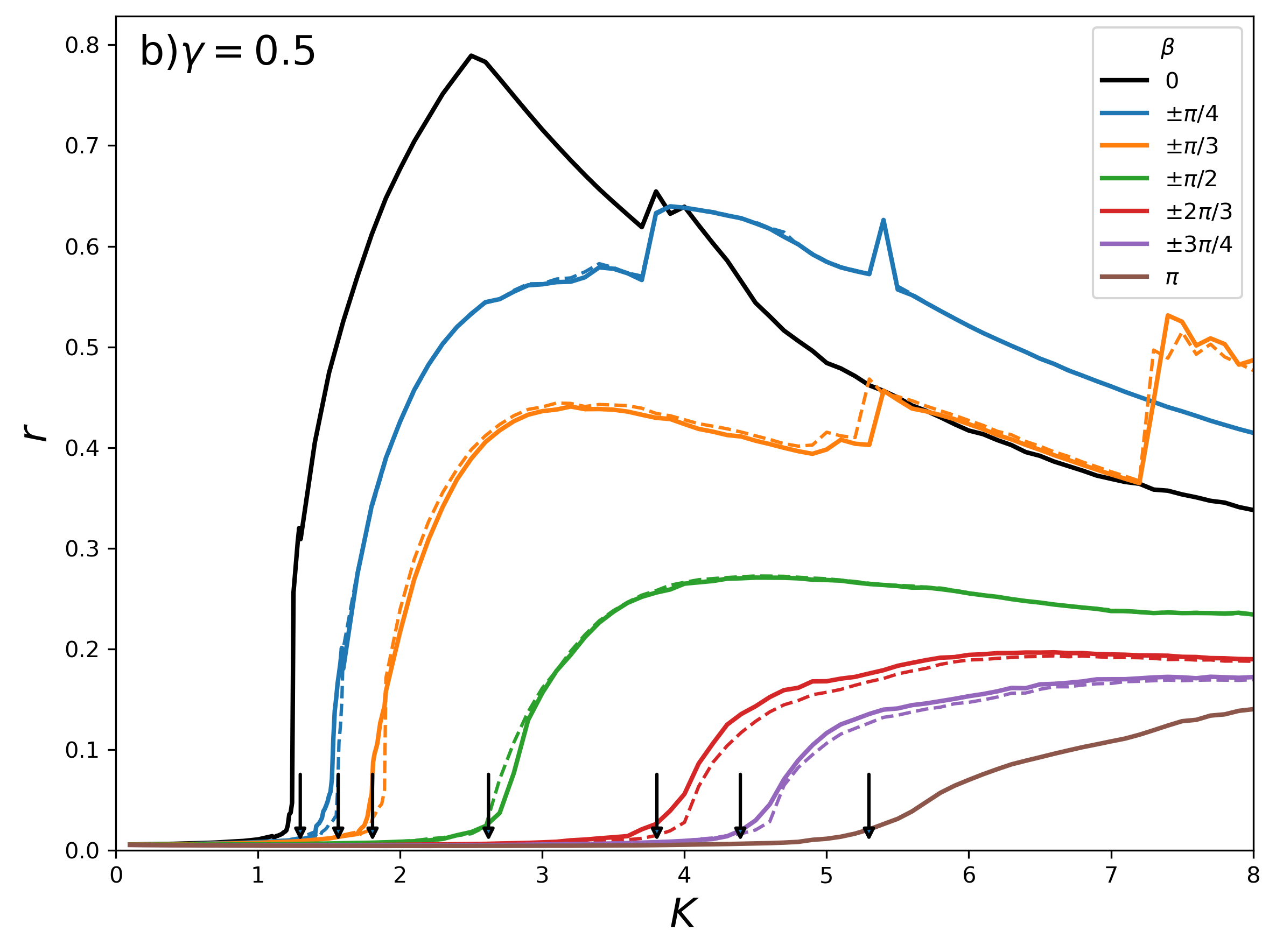}
\includegraphics[width=\columnwidth]{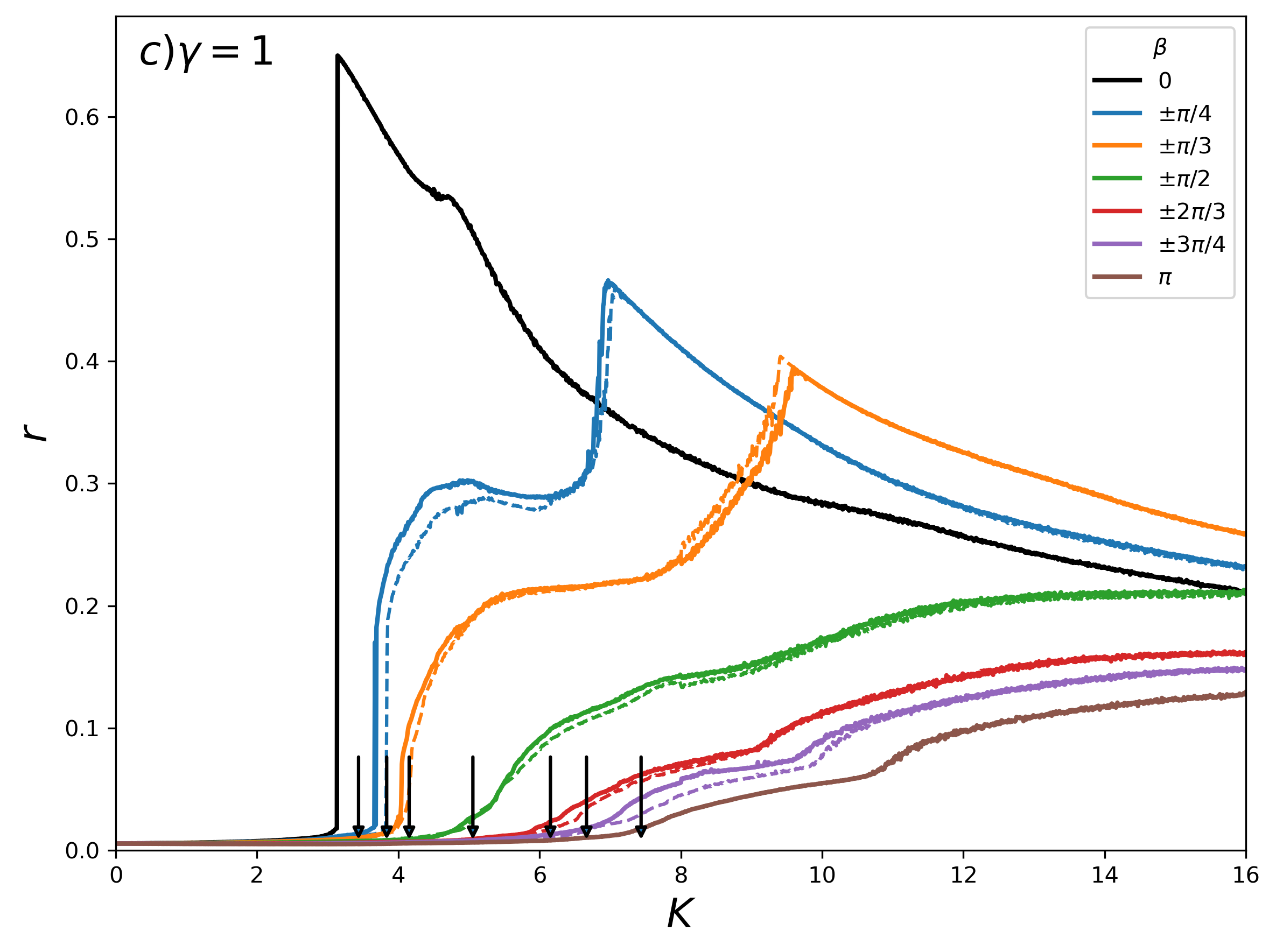}
\caption{Time-averaged order parameter $r=\langle r_n\rangle$ versus coupling $K$ for $\beta$ values as indicated in the plot; the data are obtained from direct numerical simulation of the dynamics~\eqref{eq:EOM}, after discarding $1.5 \times 10^4$ time steps of transients and then averaging over $10^4$ time steps. Solid lines refer to $+\beta$, while dashed lines correspond to $-\beta$. Arrows indicate the analytical synchronization threshold in Eq.~\eqref{eq:kc-final} for  a) $\gamma=0.1$, b) $\gamma=0.5$, c) $\gamma=1$. The system size is $N=2.5\times 10^4$.}
\label{fig:gamma}
\end{figure}

\begin{figure}
\includegraphics[width=\columnwidth]{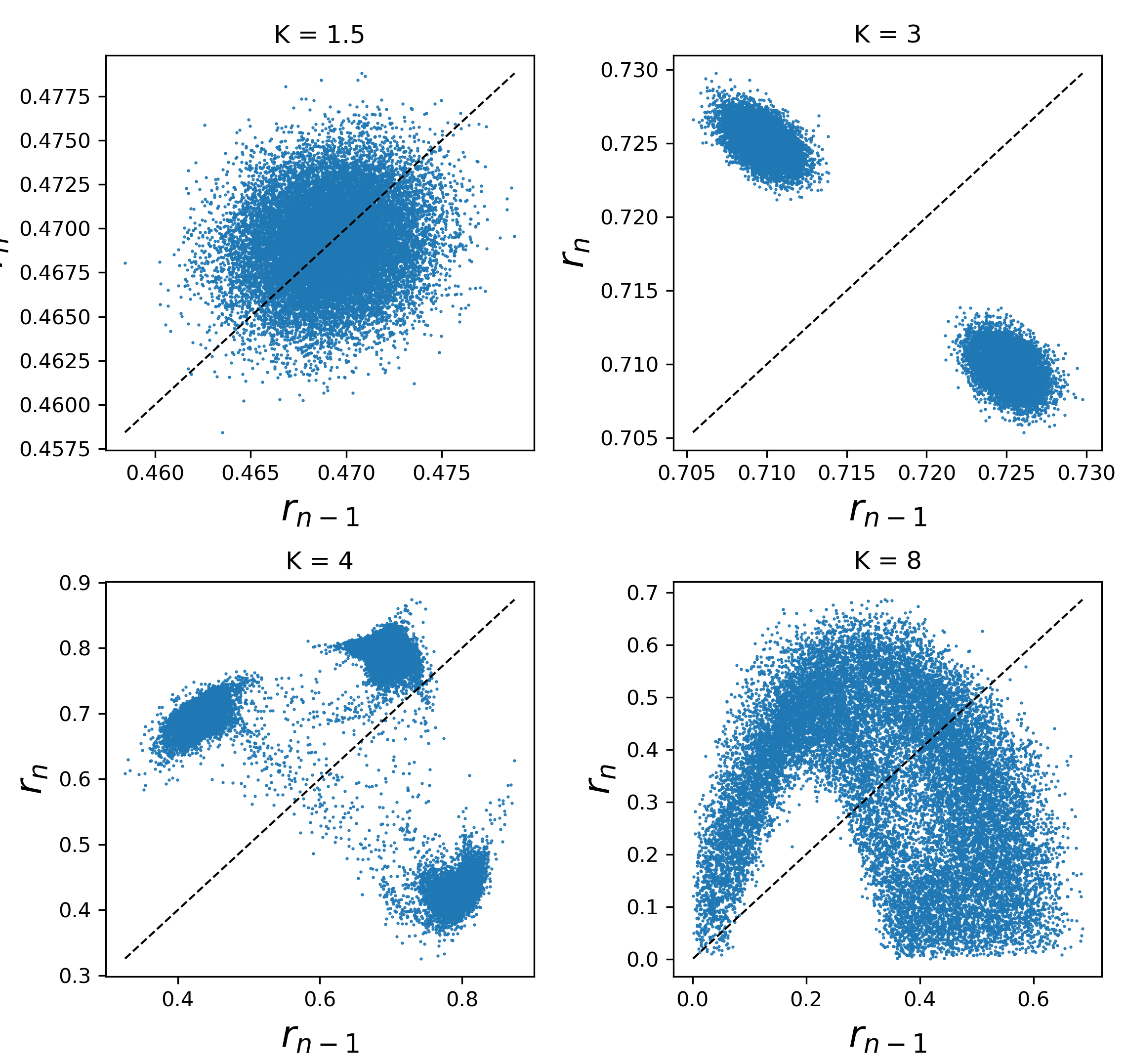}
\caption{
Return maps of the order parameter magnitude, showing $r_n$ versus $r_{n-1}$ for $\beta=0$, $\gamma=0.5$, $N=2.5\times 10^4$; here, $K_c \approx 1.297$. The panels correspond to (a) $K=1.5$, (b) $K=3$, (c) $K=4$, and (d) $K=8$. The dashed line stands for $r_n=r_{n-1}$. The data are obtained from direct numerical simulation of the dynamics~\eqref{eq:EOM}, and correspond to discarding $1.5\times 10^4$ time steps of transients.}
\label{fig:return}
\end{figure}

\begin{figure*}[t]
\centering
\includegraphics[width=\textwidth]{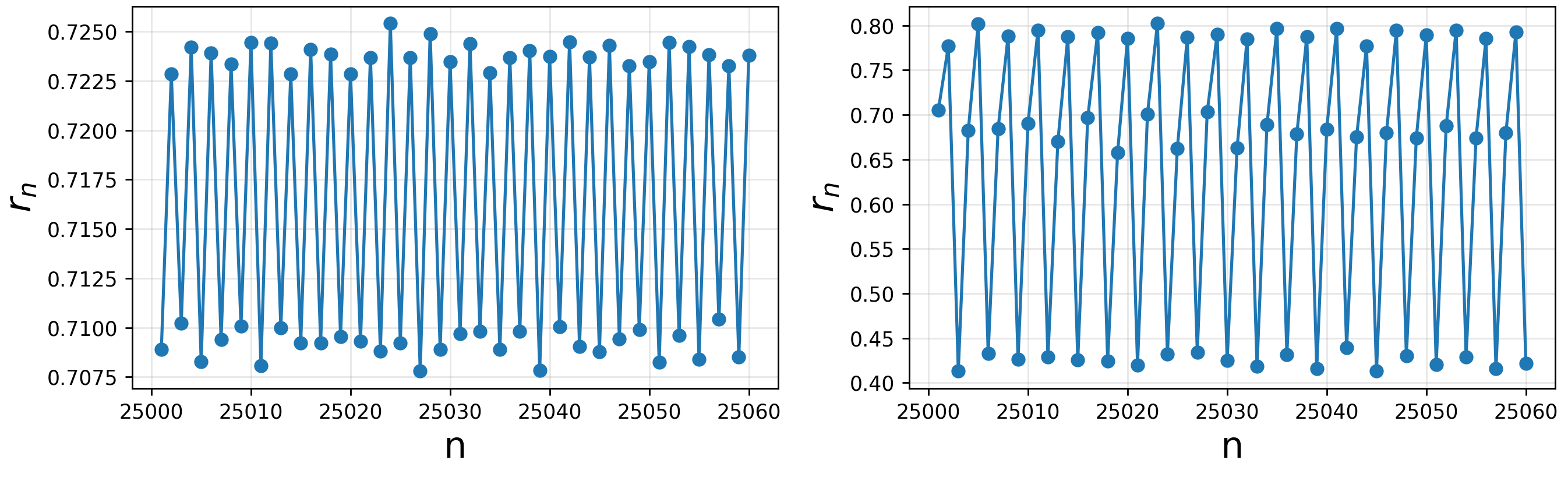}
\caption{Numerical simulation data for the time series of the order parameter magnitude $r_n$ for $\beta=0$, 
$\gamma=0.5$, $N=5\times 10^4$,  after discarding $2.5 \times 10^4$ time steps of transients. (a) $K=3.0$: implies a period-two orbit, 
(b) $K=4.0$: implies a period-three orbit,}
\label{fig:rt}
\end{figure*}

\section{Comparison between analytical and numerical results}
Figure~\ref{fig:gamma} shows numerical simulation data for the average  parameter $r=\langle r_n\rangle$ as a function of $K$ (here, $\langle\cdot\rangle$ is the time average of
$r_n$ over $10^4$ steps, after discarding $1.5 \times 10^4$ time steps of transients.).
The arrows mark the analytical threshold $K_c$ from Eq.~\eqref{eq:kc-final}; one may observe a very good agreement with the point of onset of synchronization as seen in simulations for $\gamma=0.1$ and $\gamma=0.5$. For $\gamma=1$, the match is reasonably good. Several notable features emerge from the numerical results. First, as predicted by the threshold formula, the threshold is symmetric under $\beta \to -\beta$. More surprising fact is that the entire curves for $\beta = \pm \pi/4$, $\pm \pi/3$, $\pm \pi/2$, $\pm 2\pi/3$ and $\pm 3\pi/4$ almost coincide, especially for smaller $\gamma$. Second, the threshold increases monotonically with $|\beta|$, reaching a finite value for $\beta = \pi/2$. This behavior differs markedly from the continuous-time Kuramoto model, where synchronization is completely suppressed for $\beta = \pi/2$ regardless of the coupling strength, see Eq.~\eqref{eq:Kc-continuous}. Even for $\beta=\pi$, which is a clear repulsive interaction, there is synchronization transition, as may be seen from the figure. Third, above the transition, the order parameter exhibits non-monotonic behavior for small $\beta$: a cusp-like maximum is followed by a decrease in $r$ as $K$ is increased further, indicating that stronger coupling can actually reduce synchrony in the discrete-time system. Finally, for larger values of $K$, the dynamics of the order parameter becomes increasingly complex, as we now discuss.  

For $\beta=0$ and $\gamma=0.5$, return maps of $r_{n}$ versus $r_{n-1} $ are shown for four
values of $K$ in Fig.~\ref{fig:return}.  At $K=1.5$, the dynamics fluctuates around a single synchronized state, producing a compact cluster near the diagonal. At $K=3$, the return map collapses onto two clusters, indicating a period-two orbit of the macroscopic order parameter. At $K=4$, three distinct clusters are observed, corresponding to a period-three orbit. For $K=8$, the return map develops a stretched, folded structure reminiscent of a horseshoe map, suggesting the onset of chaotic macroscopic dynamics. At $K=3$, the return map collapses into two distinct clusters, while at $K=4$, three distinct clusters emerge. At $K=8$, the return map develops a stretched and folded structure reminiscent of a horseshoe map, indicating more complex dynamics.  These observations demonstrate that increasing the coupling strength leads to a sequence of increasingly complex collective states that lie beyond the regime captured by the linear stability analysis.

The coarse-grained period-two and period-three orbits are clearly manifest in the time series of 
$r_n$ for $N=5\times 10^4$, recorded after discarding $2.5 \times 10^4$ time steps of transients, see Fig.~\ref{fig:rt}.
At $K=3$, $r_n$ alternates between two well-separated values,
$\approx0.71$ and $\approx 0.722$, corresponding to the two clusters of the return map
in Fig.~\ref{fig:return}b). Increasing the system size  reduces
finite-size fluctuations, and periodic orbits are 
sharply resolved. At $K=4$, $r_n$ cycles through three distinct
levels: a low state with a value in the range $0.43$--$0.45$, and two high states with values 
in the range $0.68$--$0.72$ and in the range $0.78$--$0.82$, manifesting the three clusters in
Fig. \ref{fig:return}c).

\section{Remarks on the Ott-Antonsen (OA) ansatz}
\begin{figure}
\includegraphics[width=\columnwidth]{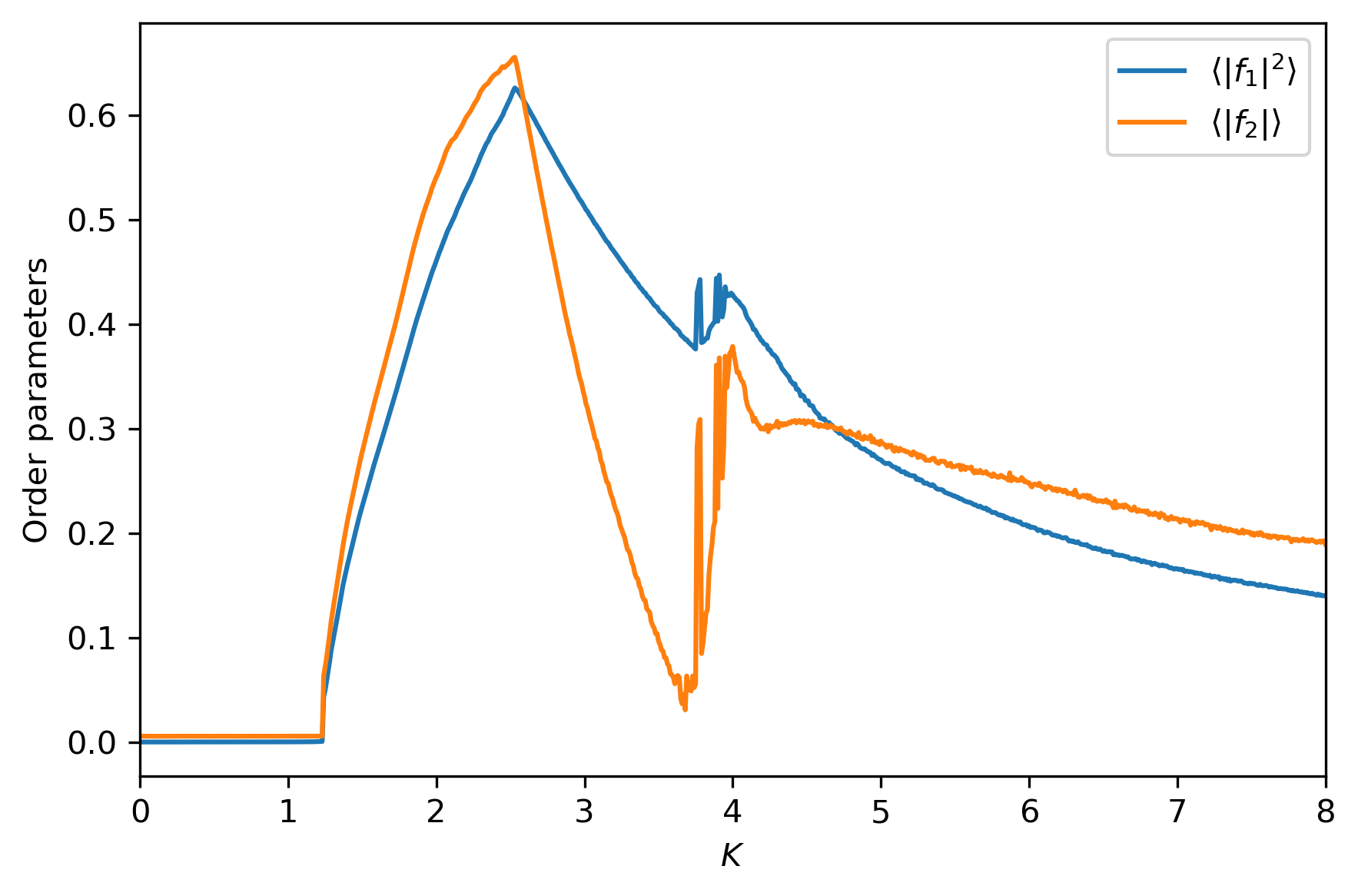}
\caption{
Comparison of the first and second Fourier modes of the phase distribution for $\beta=0$, $\gamma=0.5$, $N=2.5\times 10^4$. Shown are the quantities $\langle |f_1|^2\rangle$ and $\langle |f_2|\rangle$ as a function of $K$, obtained from direct simulation of the dynamics~\eqref{eq:EOM}, after discarding $1.5\times 10^4$ time steps of transients. }
\label{fig:f1f2}
\end{figure}

To assess the validity of the OA ansatz, we measured the magnitude of the first and second Fourier modes of the phase distribution from direct numerical simulations, as
$f_m=(1/N)\sum_{j=1}^{N}e^{im\theta_j}$. On the OA manifold, the Fourier coefficients satisfy the relation $f_m=(f_1)^m$, implying $|f_2|=|f_1|^2$. Figure~\ref{fig:f1f2} compares the numerically-measured values of the quantities $\langle |f_1|^2\rangle$ and $\langle |f_2|\rangle$ as a function of the coupling strength $K$. While the two quantities are nearly identical close to the synchronization threshold $K_c=2(e^\gamma-1) \approx 1.297$, clear deviations develop deep in the synchronized regime, i.e., for $K>K_c$, see Fig.~\ref{fig:f1f2}. This is in contrast to the continuous-time case, where the OA manifold is invariant and provides an exact low-dimensional reduction of the dynamics. We emphasize that the analyticity assumption used in the linear stability analysis presented in this work is a much weaker condition than the full OA ansatz. It only requires that the first Fourier mode $\hat f_1(\omega,n)$ 
is analytic in the lower-half complex-$\omega$ plane, which is preserved by the linearized 
dynamics. The full OA ansatz, by contrast, imposes a specific algebraic 
structure on all Fourier modes simultaneously, which as discussed above is not preserved 
by the discrete-time dynamics.

\section{Conclusion}
We have studied the discrete-time Kuramoto model with phase lag and determined the stability threshold of the incoherent state. In the continuum limit, the dynamics is governed by the Frobenius-Perron equation for the single-oscillator density. Linearization around the incoherent state leads to the stability threshold
$K_c = 2(\sqrt{e^{2\gamma} - \sin^2\beta} - \cos\beta)$, which is verified numerically. 
Remarkably, synchronization persists even for purely repulsive coupling:
synchronized states with $r>0$ are observed well above $\beta>\pi/2$, in contrast to the continuous-time Kuramoto model where no synchrony is present in such a regime.
We have also examined the applicability of the OA ansatz, and have found that the standard OA description does not provide an accurate low-dimensional description. This underscores a fundamental 
difference in synchronization transition of flows (continuous-time dynamics) and maps (discrete-time dynamics). In contrast to the classical continuous-time Kuramoto model, where the
Ott-Antonsen manifold for Lorentzian $g(\omega)$ yields a one-dimensional
dynamics for $r$ that relaxes to a fixed point and forbids periodic or
chaotic macroscopic states, our discrete-time model admits such possibilities.  Future work could explore the possibility of chimera states in discrete-time systems, and extensions to the Kuramoto-Sakaguchi model with nonuniform coupling. Our results provide a benchmark for studies 
of synchronization in iterated maps with phase delays.

\section*{Data Availability}
The data that support the findings of this study are available from the corresponding author upon reasonable request.

\acknowledgments
This work was supported by the Department of Atomic Energy, Government of India, under Project Identification Number RTI-4012. The computations were carried out on the computing clusters at the Department of Theoretical Physics, TIFR, Mumbai. We also thank Ajay Salve and Kapil Ghadiali for their computational support. 

\bibliography{References}

\end{document}